\documentstyle[prl,aps,epsfig,floats]{revtex}
\def\DESepsf(#1 width #2){\epsfxsize=#2 \epsfbox{#1}}
\begin{document}
\pagestyle{empty}                                      
\preprint{
\hbox to \hsize{
\hbox{
            }
\hfill $
\vtop{
 \hbox{ }}$
}
}
\draft
\vfill
\twocolumn[\hsize\textwidth\columnwidth\hsize\csname 
@twocolumnfalse\endcsname
\title{\boldmath{
Probing for the Charm Content of $B$ and $\Upsilon$ Mesons}
      }
\vfill
\author{$^{1}$Chia-Hung V. Chang and $^2$Wei-Shu Hou}
\address{
\rm $^1$Department of Physics, National Taiwan Normal University,
Taipei, Taiwan, R.O.C.\\
\rm $^2$Department of Physics, National Taiwan University,
Taipei, Taiwan, R.O.C.
}

%
%
\vfill
\maketitle
\begin{abstract}
A slow $J/\psi$ bump exists in 
the inclusive $B\to J/\psi + X$ spectrum, while the 
softness of $J/\psi$ spectrum in $\Upsilon(1S) \to J/\psi + X$ decay 
is in strong contrast with expectations from color octet mechanism.
We propose {\it intrinsic} charm as the explanation:
the former is due to $\bar B\to J/\psi D \pi$,
with three charm quarks in the final state;
the latter is just a small fraction of
$\Upsilon(1S) \to (c\bar c)_{\rm slow} + 2\;$``jet" events,
where the slow moving $c\bar c$ system evolves into $D^{(*)}$ pairs.
Experimental search for these phenomena
at B Factories and the Tevatron is strongly urged,
as the implications go beyond QCD.
\end{abstract}
\vskip2pc]

\pagestyle{plain}

Owing to its heaviness and narrow width,
the $J/\psi$ meson has helped shed much light on
the underpinnings of Quantum Chromodynamics (QCD),
the more recent example being the inclusive production
of energetic $J/\psi$'s in various processes.
Thus, the inclusive spectrum of 
$B\to J/\psi + X$ decay is largely understood, 
{\it except for a slow $J/\psi$ bump} 
observed \cite{CLEO} by CLEO and  recently
confirmed \cite{Belle} by the Belle Collaboration
 (see Fig.~1(a)).
Another intriguing old result from CLEO
could be related, namely
the peaking of $p_{J/\psi}$ below 2 GeV for 
inclusive $\Upsilon(1S) \to J/\psi + X$ decay \cite{CLEO1S},
which is in strong contrast (see Fig.~1(b)) with 
the hard spectrum expected from color octet mechanism.
In this Letter we attempt at linking these two
soft $J/\psi$ production phenomena
and propose specific, testable
mechanisms.

\begin{figure}[b!]
\vskip-0.3cm
\centerline{\DESepsf(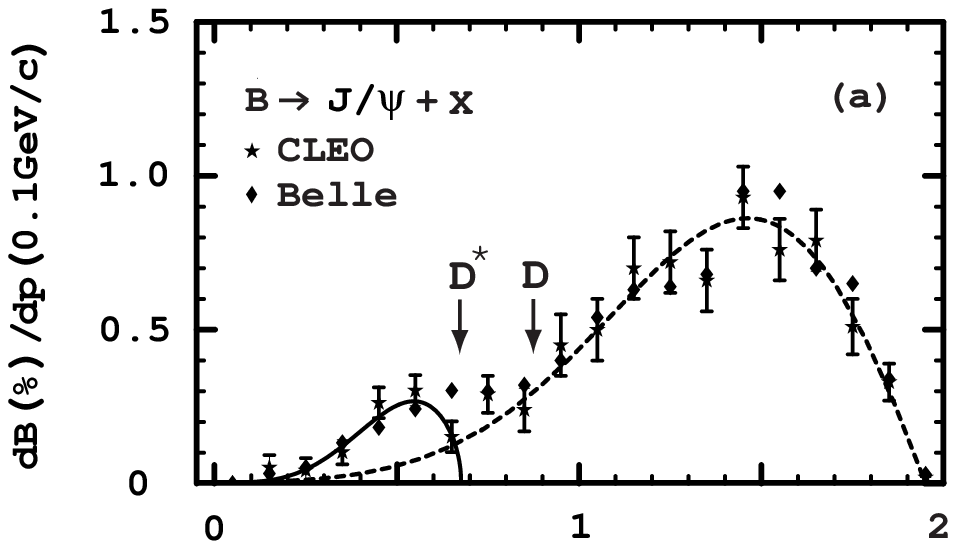 width 7.0cm)}
\vskip-0.4cm
\centerline{\DESepsf(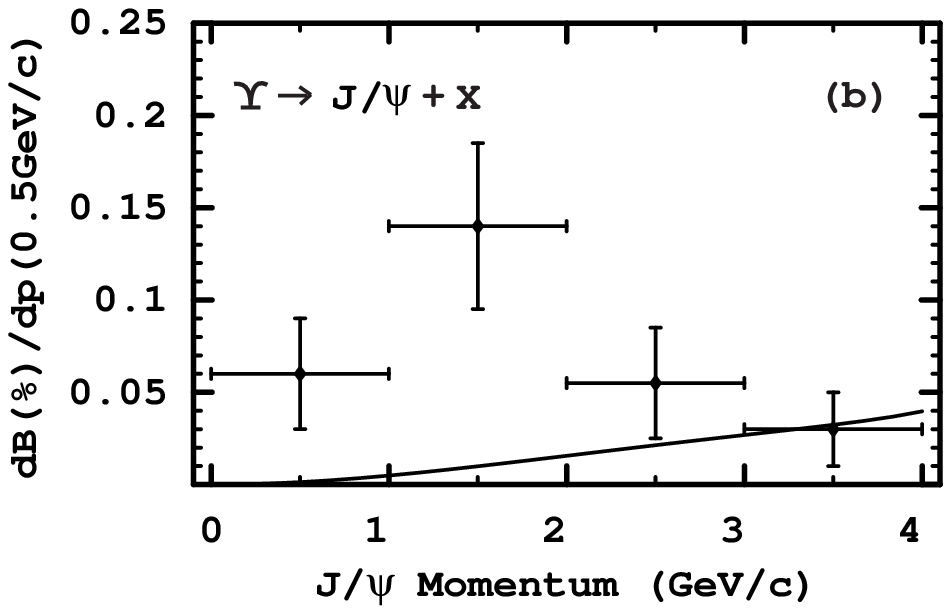 width 7.0cm)}
\smallskip\smallskip\smallskip
\caption{
(a) Inclusive $B\to J/\psi +X$ spectrum with 
$B\to \psi^\prime +X$ and $\chi_c +X$ feed-down subtracted. 
The stars (diamonds) with(out) errors are from Ref.~[1] ([2]).
The dashed curve is a simple modified phase space fit.
The solid line fits below $D^*$ threshold 
assuming recoiling $D\pi$;
$D$ recoil is also indicated.
(b) Inclusive $\Upsilon(1S)\to J/\psi +X$ spectrum from CLEO~[3].
The curve is from color octet mechanism.}
\end{figure}

We suggest that {\it intrinsic} charm (IC) content 
may be better probed in {\it heavy} hadrons such as $B$ mesons.
Intrinsic, in contrast to extrinsic, charm content of hadrons~\cite{IC}
carry large momentum fraction
and are in the lowest energy configurations
as they arise from energy fluctuations.
Thus, in the decay of heavy mesons,
the production of slow $J/\psi$ and $\eta_c$ mesons is favored.
From the indication that IC in proton could be $\sim$ 1\% \cite{HSV},
we find that IC in $B$ could lead to $\bar B\to J/\psi D \pi$
 ({\it i.e. $c\bar cc$ in $\bar B$ decay final state!})
at the few $\times 10^{-4}$ level and 
can in principle account for the above excess.
From $\Upsilon(1S) \to J/\psi\vert_{\rm soft} + X$  we argue that
$\Upsilon(1S) \to D^{(*)}\vert_{\rm slow} + X$ 
could be quite abundant.
These modes are rather promising at the B Factories
(the $\Upsilon(1S)$ especially at CLEO) and at the Tevatron,
the observation of which would provide strong impetus
for establishing intrinsic charm in hadrons.
In turn, this would have implications for the flavor program,
such as the extraction of $V_{cb}$ from $B\to D^*\ell\nu$.

Using 1.12 fb$^{-1}$ data on $\Upsilon(4S)$ resonance,
CLEO has published 
the inclusive $J/\psi$ spectrum from $B$ decays~\cite{CLEO}.
The result, with feed-down from 
$B\to \psi^\prime +X$ ($\psi^\prime\to J/\psi\pi\pi$) 
and $\chi_c +X$ ($\chi_c\to J/\psi\gamma$) subtracted,
is reproduced as stars with errors in Fig. 1(a). 
Evidently there is some activity around $p_{J/\psi} \sim$ 0.4--0.6 GeV.
In 2000, Belle gave preliminary results \cite{Belle} 
on inclusive $B\to J/\psi +X$ 
based on 6.2~fb$^{-1}$ data.
Subtracting the $B\to \psi^\prime +X$ and $\chi_c +X$
feed-down by ourselves, we plot in Fig. 1(a) the result as 
diamonds without errors, which
confirms that there is some activity below 0.8 GeV.

%
%
%
%

To bring home the point, some modeling of 
the inclusive spectrum has to be taken.
For simplicity, we adopt the 
``modified phase space" approach \cite{BN}
and modulate a constant matrix element squared by
\begin{equation}
f(p) = p(p_{\rm max}-p)/p_{\rm max}^2 \,
     \, e^{-(p-p_0)^2/\sigma_0^2},
\end{equation}
where $p_{\rm max} = 1.95$ GeV is the maximum $J/\psi$ momentum,
and $p_0$ and $\sigma_0$ are adjustable parameters.
Taking a simple average (not plotted) of CLEO and Belle data, 
we adjust $p_0$, $\sigma_0$ to 1.9, 0.8 GeV 
and give the dashed line in Fig.~1(a) as a plausible fit.
The apparent excess in 0.3 GeV $\lesssim p_{J/\psi} \lesssim$ 0.8~GeV
is of order $5\times 10^{-4}$,
comparable to the rate for $B\to J/\psi K_S$. 

We stress that more sophisticated models 
do not change the above result.
For example, a Fermi motion model with 
$p_F \simeq 0.57$~GeV for $b$ quark inside the $B$ meson
can give a good fit \cite{PPS} to data above 0.8 GeV.
A parton-based model with soft $b$ quark momentum can also
work, albeit less well.
In both cases one invokes NRQCD and the color octet mechanism,
but one is unable to fit the low $p_{J/\psi}$ excess.
In fact, these models have a softer low $p_{J/\psi}$ tail than
the simple approach of Eq.~(1),
making the excess even more striking.

The $D$ and $D^*$ recoil thresholds
at $p_{J/\psi} = $ 0.88 and 0.66 GeV are indicated in Fig.~1(a).
There is no apparent excess for $B\to J/\psi D$. 
Taking into account the broadening from 
$B$ motion in $\Upsilon(4S)$ frame, 
Belle data might indicate the presence of $B\to J/\psi D^* \sim 10^{-4}$.
Since $D^*$ marks the opening of the $D\pi$ threshold, we adapt Eq.~(1)
to $p_{\rm max}$, $p_0$, $\sigma_0 =$ 0.66, 1.4, 1.0 GeV,
and fit with the solid curve in Fig.~1(a).
It appears that {\it $B\to J/\psi D\pi \sim 4 \times 10^{-4}$ 
could account for the lump at low $p_{J/\psi}$} \cite{Dpipipi}.
With three charm quarks in the final state,
it would be rather distinct and should be searched for.
The challenge, however, is to account for such rates.

\begin{figure}[b!]
%
%
\centerline{\DESepsf(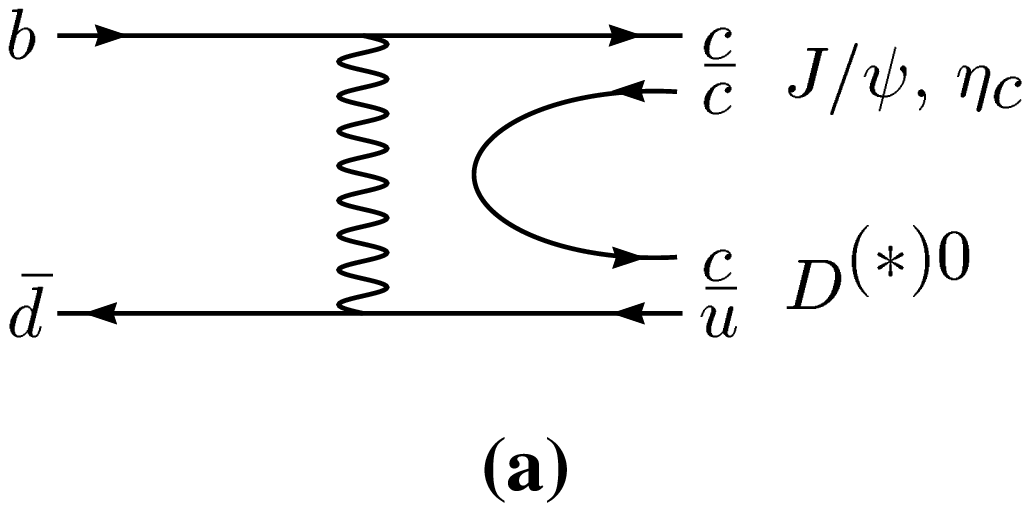 width 4.0cm) \hskip0.2cm
            \DESepsf(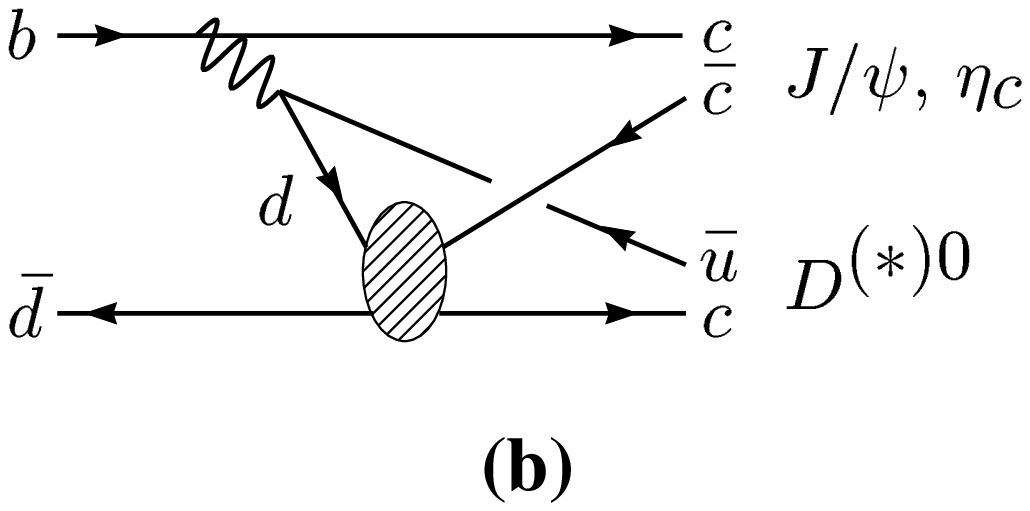 width 4.0cm)}
\vskip0.1cm
\centerline{\DESepsf(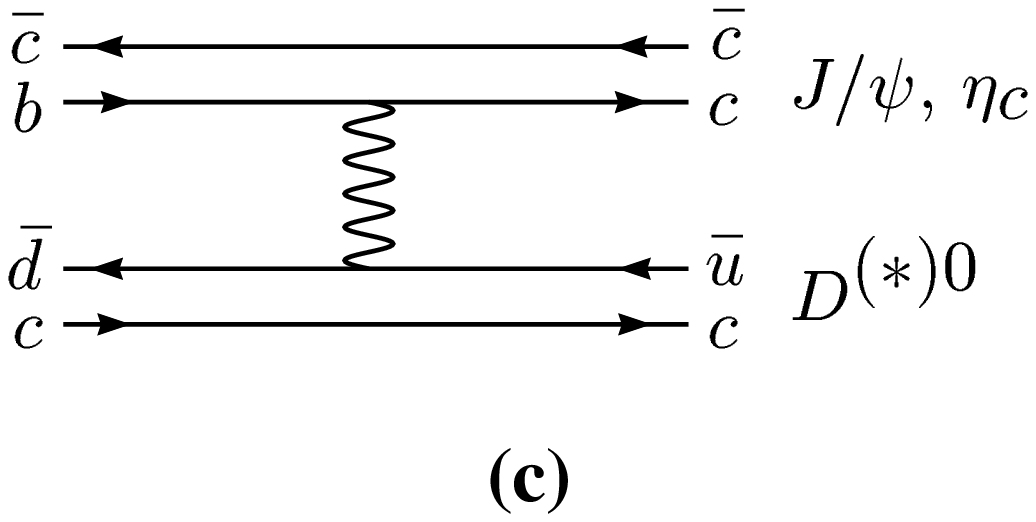 width 4.0cm) \hskip0.2cm
            \DESepsf(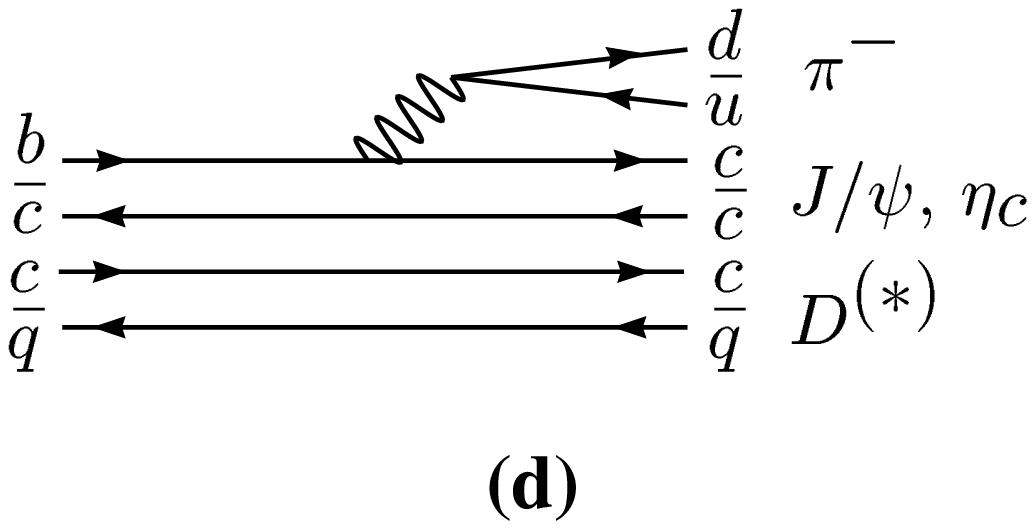 width 4.0cm)}
\vskip0.1cm
\centerline{ \hskip-0.05cm
	    \DESepsf(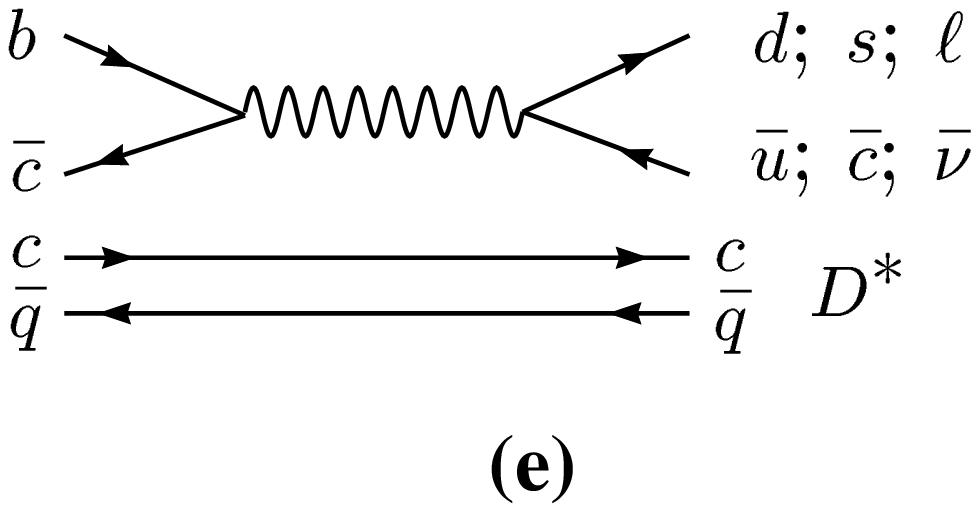 width 3.8cm) \hskip0.33cm
            \DESepsf(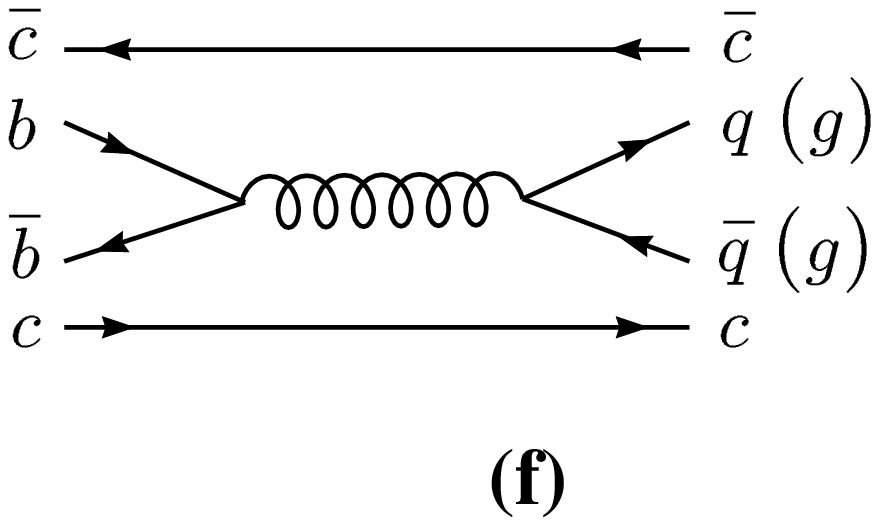 width 3.45cm) \hskip0.4cm}
\smallskip\smallskip
\caption{
Standard $\bar B^0 \to J/\psi D^{(*)0}$ via
(a) exchange and 
(b) 
    rescattering;
$\bar B \to J/\psi D^{(*)}(\pi)$ via
(c) exchange and
(d)~spectator from $\vert b\bar c c\bar q\rangle$ Fock component, and 
(e) $b\bar c$ annihilation;
(f)~$\Upsilon(1S) \to J/\psi + X$ via
 $\vert b\bar c c\bar b \rangle$ Fock component.
}
\end{figure}

Two possible diagrams are given in Figs. 2(a) and 2(b).
The first involves $W$ exchange and is of annihilation type.
Since nonperturbative $c\bar c$ production is exponentially suppressed,
the leading mechanism is perturbative and via one gluon.
Collecting factors, we estimate that the rate
should be suppressed by
$(f_B/m_B)^2\times (\alpha_s/\pi)^2\times 1/3$
compared to $\bar B\to DD_s^- \sim 1\%$, 
where $p_{J/\psi}/p_{D_s} \sim 1/3$ comes from phase space.
This gives 
$\bar B^0\to J/\psi D^{(*)0} \sim 10^{-7}$ from Fig. 2(a) alone.

Fig. 2(b) is much harder to estimate, since it involves 
$d\bar d\to \bar cc$ rescattering.
Assuming that this occurs perturbatively via one gluon, 
we estimate the rate from Fig.~2(b) alone to be
no more than $\zeta_{\bar d}\times (\alpha_s/\pi)^2\times 1/3$
times $\bar B\to DD_s^-$ rate $\sim 1\%$.
Since the rescattering demands 
energetic $d$ quark {\it and} spectator $\bar d$ quark
to produce the heavy $\bar cc$ system, 
the $\zeta_{\bar d}$ factor is expected to be considerably less than one,
hence $\bar B^0\to J/\psi D^{(*)0}$ from Fig. 2(b) 
should be safely below $10^{-5}$.
As this is not a firm result,
we note that
if one replaces $c\bar c$ by $s\bar s$ in Figs.~2(a) and 2(b), 
one already has the limit $\bar B^0 \to D_s^+ K^- \lesssim 10^{-4}$
from data \cite{DsK}.
This can be viewed as an upper bound on
$\bar B^0\to J/\psi D^{(*)0}$ modes,
where the limit can be improved at the $B$ Factories.
Similarly, rescatterings such as 
$\bar B^0 \to D^+\pi^- \to J/\psi D^0$
(corresponding to two cuts of Fig. 2(b))
should not only suffer from cancellations
between the large number of possible diagrams,
but can also be firmly bound by searching for 
color suppressed modes such as $D^0\pi^0$. 

We thus see that, 
while hard predictions of hadronic $B$ decays are difficult,
there would likely be no plausible explanation for 
$B\to J/\psi D\pi$ if it is observed
at $\times 10^{-4}$ level or higher.

It is intriguing that a soft $J/\psi$ problem exists
for $\Upsilon(1S)$ decays as well.
Based on $\sim 7\times 10^5$ $\Upsilon(1S)$ events 
and $\sim 20$ $J/\psi \to \mu^+\mu^-$ candidates, 
CLEO observed ~\cite{CLEO1S} some time ago 
that $\Upsilon(1S)\to J/\psi +X \sim 1.1\times 10^{-3}$.
The rate is still not sufficiently accounted for \cite{CKY},
but the most striking feature from data is 
the relatively soft $p_{J/\psi}$ spectrum 
that seemingly peaks below 2 GeV, as shown in Fig. 1(b),
compared to $p_{\rm max} \simeq 4.2 $ GeV.
Perturbative production predicts a hard $J/\psi$ spectrum \cite{CKY}
(solid curve).
Thus, this old CLEO result, if confirmed, 
would be a major puzzle for the color octet mechanism.

Figs. 1(a) and 1(b) together suggest 
some additional mechanism that is responsible for 
soft $J/\psi$ production from heavy meson decays.
We propose one possibility, namely {\it intrinsic charm}.
IC of hadrons in principle should exist~\cite{IC},
and has been suggested to account for 
charm production in deep inelastic scattering \cite{HSV}
and $J/\psi \to \rho\pi$ decay \cite{BK}.
It has also been suggested \cite{EFHK} 
for $D$ and $B$ mesons though never pursued.

For the proton,
$\vert p \rangle = \Psi^p_{uud} \, \vert uud \rangle
                 + \Psi^p_{u \bar c c ud} \, \vert u \bar c c ud \rangle
                 + \cdots$,
the $\vert u \bar c c ud \rangle$ component \cite{BHMT}
is generated by virtual $gg \to c\bar c$ interactions
(so multi-connected to valence quarks),
and should scale as $\alpha_s^2(m_c^2)/m_c^2$ 
relative to the $\vert uud \rangle$ component.
Since this higher Fock component arises as a quantum fluctuation,
$\Psi^p_{u \bar c c ud} \propto 1/(m_p^2 - M^2)$
where $M^2$ is the invariant mass of the fluctuation.
We show in Fig.~3(a) the distributions 
in the $\vert u \bar c c ud \rangle$ component.
One sees that IC carries {\it large} momentum fraction \cite{IC}, 
in contrast to the small $x$ tendency of 
usual ``sea" (extrinsic) quarks from gluon splitting. 
Using this feature,
there is some evidence from data that IC in proton,
$\vert \Psi^p_{u \bar c c ud} \vert^2$,
could be at $\sim 0.86\%$ level \cite{HSV}.
Such an analysis, of course, should be done
in a more consistent framework~\cite{ASTW}
and incorporate more data to be conclusive.

\begin{figure}[tb]
%
%
\centerline{\hskip0.15cm
            \DESepsf(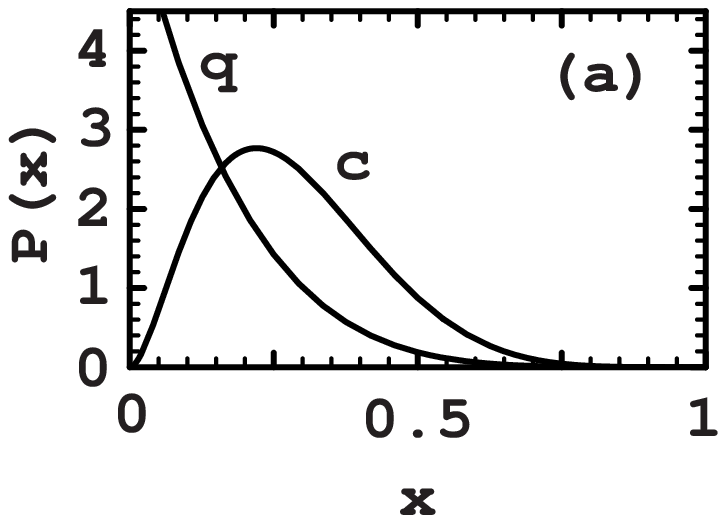 width 4.5cm) \hskip-0.45cm
            \DESepsf(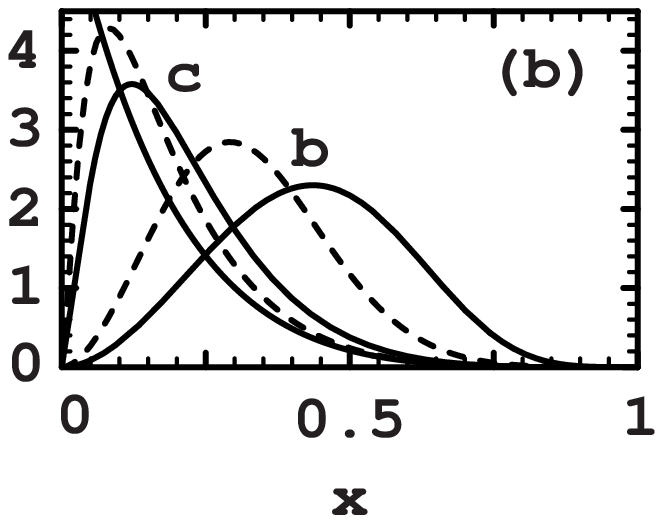 width 4.45cm)}
\smallskip\smallskip
\caption{
Intrinsic charm in 
(a) proton and (b) $B$ meson 
($b$ and $\bar q$ distributions also shown), 
with arbitrary normalization.
Dashed lines in (b) are for the 
$\vert b\bar c c\bar b\rangle$ component of $\Upsilon(1S)$.
}
\end{figure}

For $B$ mesons, 
one has the differential probability
\begin{equation}
{dP_{\rm ic}^B \over dx_1 \cdots dx_4} \propto 
\ {\alpha_s^2(m_c^2) \, \delta(1-\sum_{i=1}^4 x_i) 
                     \over
  (m_B^2 - \widehat{m}_b^2/x_b - \widehat{m}_c^2/x_c
                             - \widehat{m}_{\bar c}^2/x_{\bar c})^2},
\end{equation}
in the $\vert b\bar c c\bar q \rangle$ Fock component of
$\vert \bar B \rangle =
  \Psi_{b\bar q} \, \vert b\bar q \rangle
 +\Psi_{b\bar c c\bar q} \, \vert b\bar c c\bar q\rangle + \cdots$,
since $m_B$, $m_b$ cannot be ignored.
In fact, the heaviness of $m_b$ guarantees that it still carries
the largest momentum fraction, as can be seen from Fig.~3(b).
We find $\langle x_c\rangle \sim 0.22$ in $B$,
lower than $\langle x_c\rangle \simeq 0.28$ in
proton (Fig. 3(a)).

One has no deep inelastic scattering data off $B$ mesons
to extract $\vert \Psi_{b\bar c c\bar q} \vert^2$.
However, $\vert \Psi_{b\bar c c\bar q} \vert^2$ may be no less than
$\vert \Psi_{u \bar c c ud} \vert^2$,
because \cite{Brod} of a larger reduced mass:
the $B$ meson is more compact than usual hadrons,
hence the IC amplitude could be dynamically enhanced.
Thus,
IC in $B$ could also be $\sim$ 1\%.
A heavy quark mass expansion study \cite{FPG}
gives IC in $p \sim 10^{-3}$, which is not inconsistent.
As stressed in \cite{Brod},
this study shows that the $c\bar c$ pair
should be in the color octet configuration,
hence the IC component arises from the nonabelian nature of QCD.

To account for the low $p_{J/\psi}$ bump of Fig.~1(a)
from the $\vert b\bar c c\bar q \rangle$ Fock component,
note that $\bar B \to J/\psi D^{(*)0}$ decay via Fig. 2(c)
is still suppressed by $f_B$, just like Fig.~2(a).
The spectator decay of Fig. 2(d) is more promising.
It gives $\bar B \to J/\psi D\pi^-$ if one assumes factorization,
hence fits our interpretation of the low $p_{J/\psi}$ bump 
in Fig. 1(a) rather well. 
But can we account for the rate?

The quark level 3-body spectator decay is very sensitive to $x_b$ 
as it scales with the available energy $\Delta E$ to the fifth power.
Since $\langle x_b\rangle \simeq 0.41$ from Fig. 3(b),
taking $x_b \simeq 0.41$ to $0.6$ gives a rate in the
$10^{-3}$ to $5\times 10^{-2}$ range,
shooting up to even 10\% for $x_b = 0.65$.
We have used $\Delta E \simeq m_b - m_c \sim 3.4$ GeV 
for standard $b$ decay via $\vert b\bar q \rangle$ component, 
and $\Delta E \simeq x_b\, m_B - 1.3$~GeV
via $\vert b\bar c c\bar q \rangle$ Fock component of Fig. 2(d). 
Varying $m_c \sim 1.3$ to $1.4$ GeV does not change this range by much. 
With $x_c \lesssim \langle x_c \rangle \sim 0.22$
and $x_{\bar q}$ at its peak near zero, 
$b\to d\bar uc$ decay leads to the configuration
$(d\bar u) c \bar c (c\bar q)$,
where we have indicated color singlet pairings.
The remaining $c\bar c$ is also color singlet 
and plausibly evolves into $J/\psi$ or $\eta_c$.
With all quarks having low momenta, 
there is ample time for them to redistribute energy and momentum,
including $J/\psi D^+\pi^- \to J/\psi D^0\pi^0$.
It is sensible then that one can take the above
$\sim 10^{-2}$ factor times the IC fraction
$\vert \Psi_{b\bar c c\bar q}\vert^2$ as a rough estimate.
Note that, from duality, the ``quark level" decay rate 
can only feed into $J/\psi D^*$ or $J/\psi D\pi$ final state.
In this way, we find that a rate at few $\times 10^{-4}$ level 
is possible, {\it if} IC fraction is not much less than 1\%.
{\it One should check experimentally
whether our suggested signals are there.}

It is useful to identify other effects where 
IC of $B$ can make important impact.
Since the $b\bar c$ forms a color singlet in
$\vert b\bar c c\bar q \rangle$ Fock component,
$b\bar c \to s\bar c$, $d\bar u$ and $\ell\nu$ annihilation
can proceed without helicity suppression in the vector channel,
as illustrated in Fig. 2(e).
They in general complement standard channels 
hence would not be easy to distinguish.
However, the energy fluctuation argument suggest 
that the {\it vector} $b\bar c$ system would have near maximal mass, 
and the accompanying $c\bar q$ would end up in 
the lowest energy-momentum state available, 
that is, a slow moving $D^*$. 
Indeed, we find that $\langle x_c \rangle + \langle x_{\bar q} \rangle$
is right at the $D^*$ mass.
One therefore expects the most significant impact to be on
$B\to D^* \ell\nu$ near the zero recoil region,
where a distortion in spectrum {\it would affect 
the program of $\vert V_{cb}\vert$ extraction}. 
Thus, if IC in $B$ is truly sizable, 
there would be new systematics to $\vert V_{cb}\vert$ determination,
hence the importance is beyond QCD.

Turning to $\Upsilon(1S)\to J/\psi + X$ decay,
it is clear that IC of $\Upsilon(1S)$ 
naturally gives soft $J/\psi$ spectrum as given in Fig.~1(b).
Since the $b\bar b$ in the $\vert b\bar c c\bar b \rangle$ Fock component
should be color octet \cite{FPG},
it decays via $b\bar b \to g^* \to q\bar q$, $gg$,
as shown in Fig. 2(f) 
($b\bar b \to gg$ has an extra $t$-channel contribution),
while the accompanying $c\bar c$ is color octet
and carries minimal energy. 
Thus, one expects the underlying process
$\Upsilon(1S) \to (c\bar c)^{\rm soft}_{\rm octet} + $ 2~``jets",
where $(c\bar c)^{\rm soft}_{\rm octet}$ evolves into 
slow moving $D\bar DX$ or $J/\psi(\eta_c)X$,
in contrast with $D^{(*)}$ production
by leading particle effects from $c$-jets,
or from $g^*\to c\bar c$ splitting.

The rate is easier to estimate than the $B$ meson weak decay case.
We estimate $\Upsilon(1S) \to q\bar c c\bar q$ rate
via IC component as,
\begin{equation}
\Gamma_{q\bar c c\bar q} = 
{\Gamma_{q\bar c c\bar q} \over \Gamma_{ee}}\times \Gamma_{ee}
\sim 6 \left( {\alpha_s \over \alpha}\right)^2
 \vert \Psi_{b\bar cc\bar b} \vert^2
 \times {\rm 1.3~keV},
\end{equation}
for a single quark flavor $q$,
where 6 is from ratio of color and electric charges.
Counting 3 ($u$, $d$ and $s$) flavors and 
roughly $N_C = 3$ colors 
for $b\bar b\to gg$,
one gains an additional factor of 6.
Understandably,
{\it the process of Fig.~2(f) is rather fast!}
We should certainly demand that
$\Gamma_{q\bar c c\bar q} < 50\% \times
\Gamma_{\Upsilon(1S)}$,
which implies that the IC fraction
 $\vert \Psi_{b\bar cc\bar b} \vert^2 \lesssim 10^{-3}$.
This is not in conflict with
$\vert \Psi_{b\bar cc\bar q} \vert^2 \sim$ 1\% since 
the $\Upsilon(1S)$ is basically a nonrelativisitic bound state
hence lacks high frequency components.
But it does mean that, if IC is relevant at all,
{\it 
$\Upsilon(1S) \to (\bar c c)^{\rm slow}_{\rm octet} + q\bar q$
could easily be 10\% (or 5 keV) of $\Upsilon(1S)$ rate}.

It is remarkable that the last statement is 
consistent with all known facts.
Very few hadronic decays of $\Upsilon(1S)$ have been reconstructed so far,
while the strong $\alpha_S^3$ dependence of
$\Upsilon(1S) \to ggg$ rate certainly allows
$\Upsilon(1S) \to (\bar c c)^{\rm slow}_{\rm octet} + q\bar q \sim 10\%$.
This is also consistent with
the observed $\Upsilon(1S) \to J/\psi + X \sim 1.1\times 10^{-3}$,
since $J/\psi$ formation should be just a small fraction
of $(\bar c c)^{\rm slow}_{\rm octet}$,
while the softness of observed $J/\psi$ spectrum
is also explained.
The main part of $(\bar c c)^{\rm slow}_{\rm octet}$
evolves into $D^{(*)}\bar D^{(*)} +X$ where
the $D^{(*)}$ mesons are slow,
accompanied by two ``jets" from the $q\bar q$ and $gg$
with $5 - 6$ GeV total energy.
We note that the ARGUS Collaboration has set a bound of \cite{ARGUS}
$\Upsilon(1S) \to D^{*-}X <$ 1.9\% for $p_{D^{*}} > 0.86$ GeV.
This is not yet constraining after adjusting for $D^*$ fraction,
and in particular, one must loosen the cut on
$p_{D^{*}}$ to be sensitive to IC induced decays.
It is rather exciting that
CLEO would be running on $\Upsilon$ resonances soon \cite{cleoc},
where we may learn about relative abundance of
$\Upsilon(1S) \to (\bar c c)^{\rm slow}_{\rm octet} +$ 2 ``jet" events.
Note that $\sim 1/6 - 1/10$ of these ``jets" would be charm jets.

Some further remarks are in order.
First,
the lump below $p_{J/\psi} <$ 0.9~GeV
cannot be due to feed-down from additional
$c\bar c$ states, since
such spectrum would in general extend beyond 1 GeV,
as can be seen from data \cite{CLEO,Belle}
for $B\to \psi^\prime +X$ and $\eta_c + X$.
There is no evidence for excess above 0.9~GeV.
Second,
the low $p_{J/\psi}$ excess could 
be due to $\bar B\to J/\psi \Lambda \bar N$ \cite{BN}, 
but then a large baryon production probability is needed 
near threhold, which is not more plausible.
Third,
for intrinsic strangeness,
one can have analgous signatures such as $\bar B\to \phi D^{(*)}$, 
by making simple changes to Figs. 2(a)--(d).
While these signals should be searched for,
they are less distinct since strangeness production is less suppressed,
and multi-particle final states may be preferred
since available energy is more than 2~GeV.
An earlier suggestion \cite{EFHK} of $D^0\to \phi \bar K^0$
as evidence for intrinsic strangeness in $D$ mesons
suffers from large rescattering at $m_D$ scale.
We mention, however, the possibility of
$B\to J/\psi\, \phi K+X$, since CLEO recently observed \cite{phik}
$B\to J/\psi\, \phi K \sim 10^{-4}$.
Whether it is from intrinsic strangeness or final state rescattering,
this possibility should be studied further.
Fourth, 
analagous signatures for $B_c$ mesons and $b$-baryons
are 
$B_c \to J/\psi \, J/\psi \, (\pi)$, 
$J/\psi \, \eta_c \pi$ and 
$\Lambda_b \to J/\psi \, \Lambda_c (\pi)$.

We emphasize that the
$\bar B \to J/\psi D^+\pi^-$, $J/\psi D^0\pi^0$ 
and $J/\psi D^{*0}$ signals
can be searched for {\it right away} at
B Factories and at the Tevatron
with relative easy.
The two environments complement each other,
with excellent $\pi^0$ detection at B Factories
but larger cross section and higher boost at Tevatron,
where a large number of $J/\psi$ events have been recorded.
The $B\to D_s^{(*)}K^{(*)}$ modes should be searched for
as control on rescattering diagrams.

In conclusion,
we propose the following study plan.
B Factory and Tevatron experiments 
should scan low momentum $J/\psi$ events for 
$\bar B \to J/\psi \, D \pi$, $J/\psi\phi \bar K+X$, 
and $J/\psi \Lambda \bar N$.
If $\bar B \to J/\psi \, D \pi$ (and perhaps $J/\psi D^*$)
is established above $10^{-4}$ level,
while $D_s^{(*)}K^{(*)}$ modes are found to be far less,
one then has a good case for intrinsic charm in $B$.
CLEO should run on $\Upsilon(1S)$
and collect at least 1~fb$^{-1}$ data,
reconfirm $\Upsilon(1S) \to J/\psi(\eta_c)+X$,
and search for $\Upsilon(1S) \to D^{(*)}+X$. 
Slow moving $J/\psi$ or $D^{(*)}\bar D^{(*)}$ 
in an otherwise jetty event from $\Upsilon(1S)$ decay
would indicate,
perhaps more unequivocally, 
the existence of intrinsic charm.  
The discovery of intrinsic charm in 
$B$ and $\Upsilon$ mesons 
would not only add another twist to hadron structure, 
it would have implications on $V_{cb}$ extraction 
from $\bar B\to D^*\ell\nu$ decay as well.

This work is supported in part by
NSC 89-2112-M-002-063 and 89-2112-M-003-021, the MOE CosPA Project, 
and as part of the NCTS Topical Program on B Physics and CP Violation.
We thank Andrzej Bozek, Stan Brodsky, Tom Browder and Henryk Palka
for inspiration.


\begin{thebibliography}{99}
\bibitem{Dpipipi} $B\to J/\psi D\pi\pi$ should be extremely suppressed
by 4-body phase space near the kinematical limit.
\end{thebibliography}
\end{document}